\documentclass[runningheads]{llncs}
\usepackage[T1]{fontenc}
\usepackage{cite}
\usepackage{float}
\usepackage{graphicx}
\usepackage{booktabs}
\usepackage{cleveref}
\begin{document}
\title{Optimized User Experience for Labeling Systems for Predictive Maintenance Applications}
\titlerunning{Optimized User Experience for Labeling Systems}
%
\author{Michelle Hallmann\inst{1}\orcidID{0009-0002-9600-7575} 
\texttt{michelle.hallmann@hshl.de}
\and
Michael Stern\inst{1}\orcidID{0009-0002-7558-5384} \and
Juliane Henning \inst{1}\orcidID{0009-0004-7443-9902}
Ute Franke\inst{2}\orcidID{0009-0009-3945-4775} \and
Thomas Ostertag\inst{3}\orcidID{0009-0005-8617-3256 } \and
Joao Paulo Javidi da Costa\inst{1}\orcidID{0000-0002-8616-4924} \and
Jan-Niklas Voigt-Antons\inst{1}\orcidID{0000-0002-2786-9262}
}
 
\authorrunning{Hallmann et al.}
 
\institute{Hamm-Lippstadt University of Applied Sciences,  59557 Lippstadt, Germany \and
5micron GmbH, Rudower Ch 29, 12489 Berlin, Germany \and
OSTAKON GmbH, Forster Str. 54, 10999 Berlin}
 
\maketitle              

\noindent
This version of the contribution has been accepted for publication, after peer review. The Version of Record is available online at: https://doi.org/10.1007/978-3-031-92689-1\_13

\begin{abstract}
The maintenance of rail vehicles and infrastructure plays a critical role in reducing train delays, preventing malfunctions, and ensuring the economic efficiency of rail transportation companies. Predictive maintenance systems powered by supervised machine learning algorithms offer a promising approach by detecting potential failures before they occur, reducing unscheduled downtime, and improving operational efficiency. However, the success of such systems depends heavily on high-quality labeled data, necessitating user-centered labeling interfaces tailored to annotators’ needs for Usability and User Experience. This study introduces a cost-effective predictive maintenance system developed as part of the federally funded research project "DigiOnTrack," which combines structure-borne noise measurement methods with supervised machine learning to provide monitoring and maintenance recommendations for rail vehicles and infrastructure in rural Germany. The system integrates wireless sensor networks, distributed ledger technology for secure data transfer, and a dockerized container infrastructure hosting the labeling interface and alarming dashboard. Train drivers and workshop foreman were annotators, labeling faults on rail infrastructure and vehicles to ensure accurate predictive maintenance recommendations. The Usability and User Experience evaluation revealed that the locomotive drivers' interface achieved "Excellent Usability," while the workshop foreman's interface was rated as "Good Usability." These results highlight the system’s potential for seamless integration into daily workflows, particularly regarding labeling efficiency. However, areas such as Perspicuity require further optimization for more data-intensive scenarios. The findings offer actionable insights into the design of predictive maintenance systems and labeling user interfaces, providing a foundation for future guidelines in Industry 4.0 applications, particularly in rail transportation.

\end{abstract}

\section{Introduction}
The maintenance of rail vehicles and infrastructure significantly affects train delays and malfunctions \cite{ref_TrainDelays} and the economic success of rail transportation companies \cite{ref_Maintenance}. Delayed maintenance can result in significant failures and economic costs, while excessive maintenance without necessity leads to avoidable disruptions and expenses \cite{ref_ImplementationPM}. It becomes evident that cost- and time-effective maintenance solutions must be found for the future of the railway industry. A promising approach involves deploying affordable wireless monitoring and predictive maintenance systems for rail infrastructure and vehicles \cite{ref_DegradationPrediction}, for example, using cost-effective sensors and machine learning algorithms. These systems should adopt a systemic approach that considers the entirety of rail interactions, emphasizing the importance of data integrity and security \cite{ref_InfrastructureOptimizationHCII2024}. Predictive maintenance systems can reduce unscheduled downtime and improve safety and operational efficiency by detecting potential failures before they occur \cite{ref_PredictiveMaintenanceRailway, ref_MLPM}. One way of integrating predictive maintenance is through supervised machine learning algorithms, which require expert-assigned labels in addition to sensor data. A high-quality labeling phase is crucial to building a reliable predictive maintenance system based on supervised machine learning algorithms, as machine learning capabilities depend heavily on the quality of the training data  \cite{ref_DataQualityML, ref_LabelingSystemHCII2024}.
To minimize the annotators' workload and ensure high-quality labeled data, the labeling user interface must be tailored to the annotators' needs, demonstrating high Usability and User Experience. However, few validated guidelines exist for designing a data management system for monitoring and predictive maintenance in the rail transportation field \cite{ref_PredictiveMaintenanceRailway}, nor for designing labeling user interfaces for predictive maintenance, with even fewer specific examples presented for software solutions \cite{ref_LabelUXGuidelines}. 

This article presents a cost-effective wireless predictive maintenance system for rail vehicles and infrastructure designed for the federally funded research project "DigiOnTrack", where we assess the effectiveness of the combination of structure-borne noise measurement methods and supervised machine learning for monitoring and maintenance recommendations on rail infrastructure and vehicles in a rural area in Germany. For this purpose, two train cars and one rail section were equipped with structure-borne noise measurement sensor systems for data collection. The two sensor systems in the train cars monitored the state of the rails, while the sensor system on the rail section monitored the state of the train cars when entering and leaving the railway workshop. To apply supervised machine learning methods, we needed labels of faults in the rail infrastructure and trains in addition to the sensor data. We built a labeling user interface for this purpose, which was used by the train drivers and the workshop foreman, who served as annotators. Train drivers labeled faults on the rail infrastructure, while the workshop foreman labeled faults and train car repairs. A distributed ledger network transferred sensor and labeling data to a dockerized container infrastructure, where the labeling user interface and the alarming dashboard were hosted, and data and labels were used for data analysis and generation of maintenance recommendations. Critical for the success of our system was the design of the labeling user interface, which needed to be seamlessly integrated into our annotators' daily work routines. For this, the interface was designed following Usability guidelines and prioritizing ease of use, consistency, and quick labeling during the annotators' daily work routines. Iterative testing ensured the user interface was intuitive and required minimal training, reducing cognitive load and improving data accuracy. In the following, we present the overall system and the Usability study we conducted to evaluate the labeling system. With the presentation of the system infrastructure and the labeling user interface, we aim
to provide suggestions for the architecture of predictive maintenance systems and the design of labeling systems in Industry 4.0, specifically focusing on the rail transportation industry.

\section{Related Work}

\subsection{Cross-Industry Perspectives on Pre in recent years active maintenance}
In many industrial settings, maintenance is still predominantly carried out as preventive maintenance, which aims to prevent machine failures through scheduled maintenance cycles \cite{Levitt2011}. However, this approach poses challenges: If maintenance intervals are too short, machines are serviced more frequently than necessary, leading to unnecessary downtime and increased costs, whereas overly long intervals increase the risk of machinery breakdowns, potentially resulting in significant economic losses due to critical equipment failures \cite{ref_ImplementationPM}. With the advancing technological possibilities of Industry 4.0, condition-based maintenance has emerged, where sensors monitor the status of machines, allowing maintenance to be aligned with their actual condition \cite{Lee}. Combining condition monitoring with machine learning algorithms to generate recommended maintenance actions has led to the development of the latest concept: predictive maintenance \cite{Mobley2002}. Predictive maintenance is increasingly being adopted across various industry sectors as part of the data-driven “Industry 4.0” paradigm \cite{Achouch}. 

Various machine learning methods can be employed, with the choice of method and design of the predictive maintenance system depending on the specific industrial field, the operating environment of the machines, and the type of sensors that are used for the data acquisition \cite{zhu2024surveypredictivemaintenancesystems}. Examples are Artificial Neural Networks, decision trees, Support Vector Machines, and k-nearest Neighbors.

\subsection{Predictive Maintenance in the Rail Transportation Area}

The railway industry operates in a high-risk environment, making the safe and efficient maintenance of rail infrastructure and vehicles essential. Various years have passed, and various approaches have been explored to improve rail maintenance, rail components, and vehicles using predictive maintenance methods. Binder et al. \cite{Binder} conducted a systematic literature review and identified 24 articles investigating the use of condition monitoring systems and predictive maintenance in the railway sector. 

Among the studied components, the most researched were rails, rail joints, and elements connecting the vehicle body to the tracks. While most reviewed studies focused on fault detection, eight papers specifically addressed the use of predictive maintenance for defect prediction. To analyze the data, researchers employed various sources, including sensor data from different sensor types, operational data from railway companies, track parameters, and environmental conditions. Due to limited or poor-quality data in several cases, some authors developed physical simulation models for the components whose maintenance required improvement. 

In addition to data-related challenges, such as insufficient training data for machine learning algorithms and inadequate data quality, physics-related problems also emerged. These issues highlight the complexity of applying predictive maintenance in the railway sector.

\subsection{Predictive Maintenance System Infrastructure}

Implementing predictive maintenance systems requires robust infrastructure capable of handling complex data requirements, ensuring reliability, and maintaining operational efficiency in critical industries like railways. To address these needs, advanced systems integrate condition monitoring with additional data sources, such as structure-borne noise sensors, GPS, and environmental parameters. These systems utilize machine learning algorithms to analyze the collected data, enabling real-time fault detection and prediction \cite{ref_DegradationPrediction}.

One of the core challenges in predictive maintenance is managing data quality, as gaps in datasets or poor data quality can significantly hinder the accuracy of machine learning models \cite{ref_DataQualityML}. Data preprocessing methods like Fast Fourier Transformation (FFT) have been employed to compress data while retaining critical features, optimizing both storage and transmission for large-scale operations \cite{ref_DegradationPrediction}. 

A notable innovation in predictive maintenance system infrastructure is the application of decentralized ledger technologies (DLT). DLT provides tamper-proof data storage, ensuring data integrity and enabling secure cross-company collaboration, which is crucial in safety-critical environments like railways \cite{ref_Binder, Achouch}. This technology allows stakeholders to share and validate data seamlessly while maintaining transparency and accountability, particularly for infrastructure conditions, anomaly labeling, and maintenance activities.

Emerging best practices emphasize modular, scalable, and secure system architectures. Distributed processing and edge computing are recommended to enable real-time anomaly detection and minimize latency in decision-making processes \cite{Achouch, ref_MLFuture}. For instance, vibration-based condition monitoring has proven effective in analyzing sensor data to detect potential faults, and machine learning models such as decision trees and random forests have demonstrated success in predictive applications across various industrial fields \cite{ref_DegradationPrediction, ref_PredictiveMaintenanceRailway}. 

Finally, the success of predictive maintenance systems depends on integrating user-friendly labeling interfaces to ensure high-quality input data for machine learning algorithms. Properly designed interfaces enable domain experts to efficiently annotate data, facilitating the development of reliable predictive models \cite{ref_TediousLabeling, ref_LabelUXGuidelines}. High-quality labeled data is essential for reducing errors in prediction and enhancing system reliability, particularly in environments where operational safety is critical.

By leveraging decentralized technologies, improving data quality, and incorporating modular system designs, predictive maintenance systems can meet the demanding requirements of railway operations and similar critical industries. These advancements enable systems to scale effectively, address operational complexities, and ensure robust performance in dynamic environments \cite{ref_Binder, ref_Maintenance, ref_MLPM}.

\subsection{Usability and User Experience for Labeling Quality}
The user interface constitutes a critical feature in the design of any labeling system. The quality of the labeling highly depends on the annotating expert and their willingness to interact with the user interface. To successfully integrate the labeling user interface into the annotator’s daily working routine, designers and developers must consider Usability and User Experience. Usability aims at ensuring users achieve their goals while working with the system through simplicity, intuition, and a user-friendly design \cite{ref_UsabilityUX}. Jakob Nielson \cite{ref_UsabilityEngineering} has developed 10 heuristics to help systems achieve better Usability. These heuristics include, among other aspects, continuously communicating the system’s status in a language familiar to the user, maintaining consistency and adhering to established standards to avoid confusion, and displaying only necessary elements through aesthetic and minimal design. These Usability aspects aid developers and designers in fostering a positive User Experience. Jacobsen and Meyer, in their 2019 approach, describe User Experience as the overall experience that the user has before, during, and after using the system. This experience should ideally leave the user with positive feelings, encouraging them to return to the system \cite{ref_UsabilityUX}. Therefore, a thorough understanding of User Experience involves consideration of the user’s perception and emotions during use, as well as their needs and the degree to which these needs are fulfilled \cite{ref_UsabilityUXDesign} \cite{ref_InteraktiveSysteme}. Since the „task of labeling data is notoriously challenging and time-consuming “ \cite{Hinkle} and can be seen as tedious by some, it is crucial to design the interface in ways that facilitate gaining a better understanding of the data annotators are working with, for example through visualizations \cite{ref_TediousLabeling}. 

Our research shows few guidelines exist for designing user interfaces for labeling systems. Driving on Nielson’s \cite{ref_UsabilityEngineering}, Passos et al. \cite{ref_LabelUXGuidelines} have developed guidelines to support software engineers in designing data labeling systems. Since these guidelines have not yet been scientifically validated at the time of developing our software, we have decided to focus on the work of Jakob Nielson and use his heuristics approach as a base for our interface design.

\section{Software Design}
\subsection{Backend Infrastructure}

The backend infrastructure was designed to support a scalable and secure predictive maintenance system, leveraging advanced technologies to ensure data integrity, reliability, and efficient processing \cite{ref_Maintenance}. The distributed ledger technology (DLT) implemented across the system provided a tamper-proof method for storing and sharing data, which is essential in the railway domain where accurate and trustworthy information can prevent critical failures and delays \cite{ref_Maintenance}. This decentralized approach enabled secure cross-company data sharing while maintaining confidentiality, ensuring that sensitive sensor data could be utilized collaboratively without compromising security.

The infrastructure employed a modular and containerized architecture using Docker, which streamlined the deployment and management of various system components \cite{ref_Docker}. This included data routing, user authentication, database management, and machine learning integration for predictive analytics. The use of Docker containers allowed for efficient resource utilization and simplified scalability, enabling the system to handle large volumes of sensor data.

A key backend component was the database infrastructure, which utilized PostgreSQL for secure and efficient data storage \cite{ref_DataQualityML}. Sensor data, labeled anomalies, and machine learning outputs were stored in dedicated tables, ensuring organized and reliable access for data analysis. The infrastructure also included preprocessing mechanisms, such as Fast Fourier Transform (FFT), to reduce data size while preserving critical features, facilitating real-time transmission and storage of structure-borne noise and other sensor data \cite{ref_FFT}.

The labeling user interface, integrated into the backend via a reverse proxy approach, enhanced the accuracy of machine learning models. It enabled field experts to efficiently tag anomalies in infrastructure and vehicles, producing high-quality labeled data essential for supervised learning algorithms. Designed to integrate seamlessly into the workflow of railway professionals seamlessly, the system minimized disruption while maximizing the quality and reliability of input data.

The backend infrastructure supported the predictive maintenance system's scalability, security, and real-time performance goals by combining distributed ledger technology, containerized architecture, and robust data management strategies. This architecture ensured that the system could adapt to the railway industry's evolving needs while maintaining high data integrity and privacy standards.

\subsection{Labeling User Interface (Front End)}

\subsubsection{Labeling Use Case Development}
We developed the labeling processes in collaboration with the railroad company, the sensor system developers, and the data exchange company of our project consortium. Through meetings with train company employees and the workshop foreman, we gained insights into the maintenance procedures of the railway and the locomotives. 
Due to their technical expertise, we assigned rail fault labeling to train drivers and the train car fault labeling to the workshop foreman. The workshop foreman, who manages maintenance tasks on a computer, confirmed he could use a web-based interface for labeling between maintenance assignments. Train drivers, equipped with service tablets, could label infrastructure events at the end of their shifts. A website serves as the labeling user interface, directing users to tailored labeling dashboards after login.

\subsubsection{Design Methodology}
The labeling user interface was designed to formalize functional and non-functional user interface requirements in collaboration with stakeholders. Wireframes defined the layout, and a digital prototype was first created for iterative testing. The system was developed using Vue.js and Vuetify, emphasizing a minimalist, user-friendly design \cite{ref_MinimalismUI, ref_Vuetify}. Feedback from university design experts guided iterative improvements, resulting in the final design.

\subsubsection{Dashboard for Train Car Faults Labeling}

\begin{figure}
    \centering
    \includegraphics[width=1\linewidth]{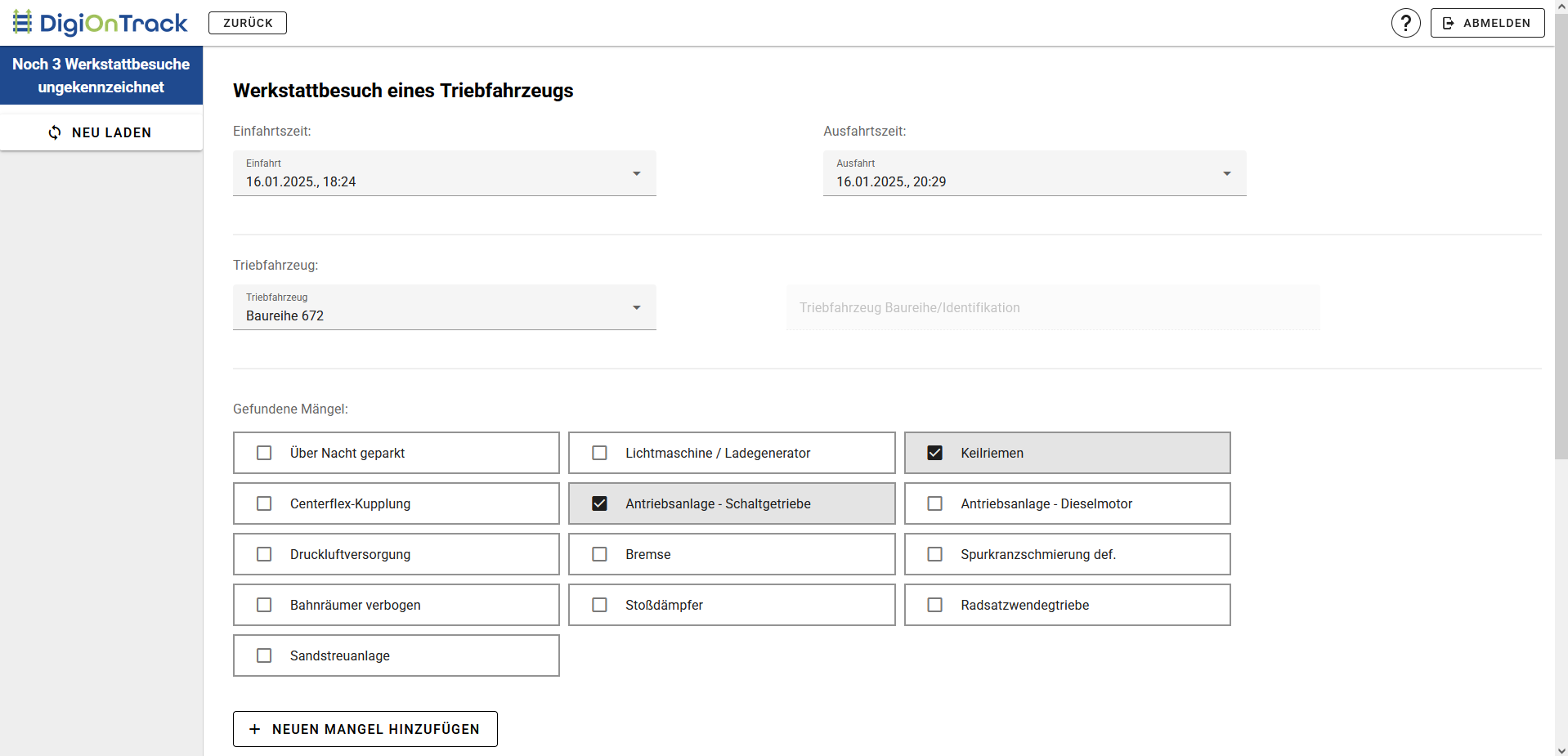}
    \caption{Dashboard for Train Car Faults Labeling of the Workshop Foreman: Back, Help, and Logout Buttons in the Header; Number of Unlabeled Workshop Visits on the Left; Entry, Exit, and Train Car Identification Selection Input Fields at the Top; Lists with Labels for Labeling of Faults and Repairs in the Center and Bottom, Button for the Creation of a New Label Below the Labels}
    \label{fig:dashboard-trainCarLabeling}
\end{figure}

\begin{figure}
    \centering
    \includegraphics[width=1\linewidth]{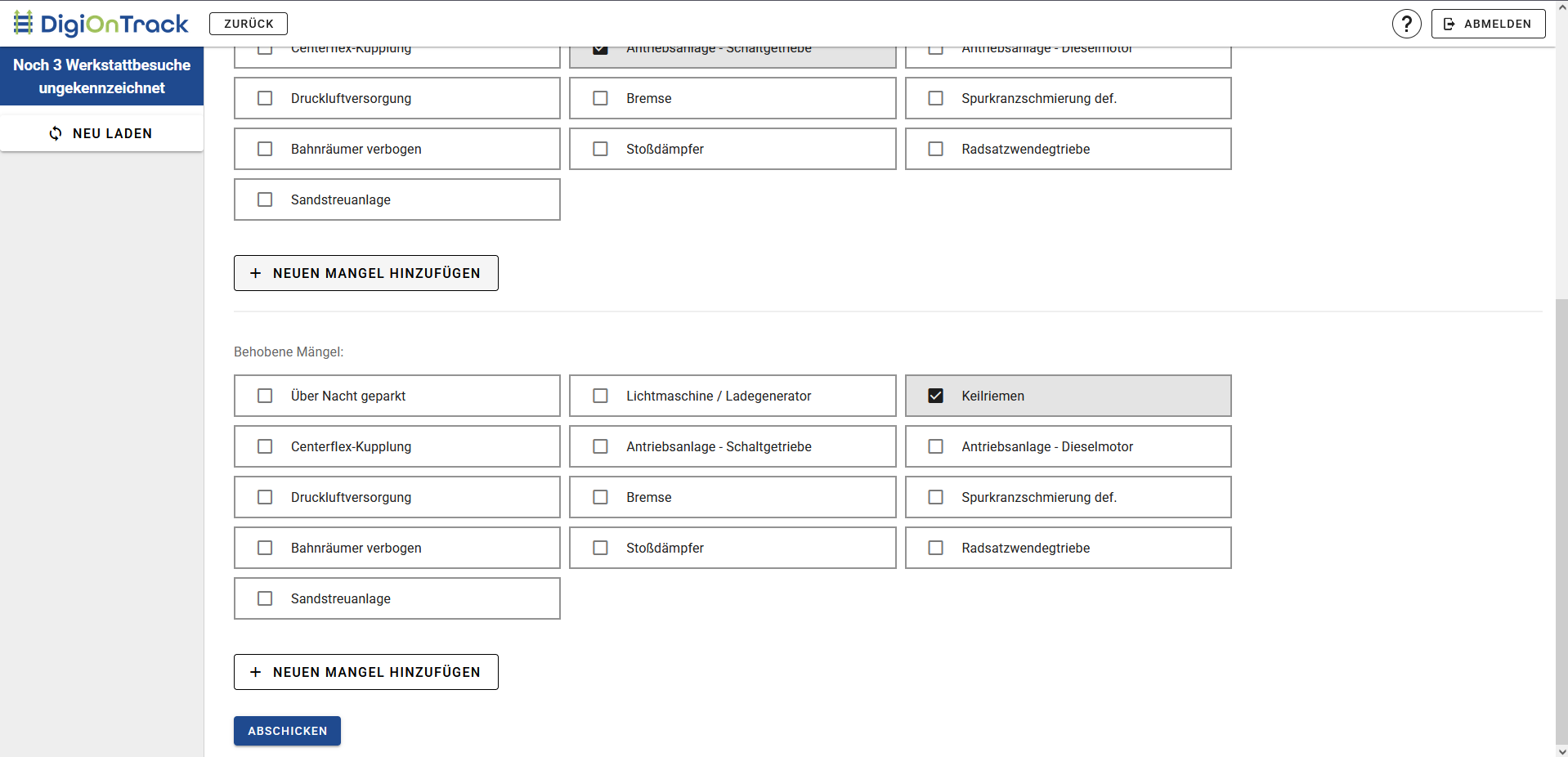}
    \caption{Dashboard for Train Car Faults Labeling of the Workshop Foreman - Continuation: Both Label Category Lists are Visible, Submit Button at the Bottom}
    \label{fig:dashboard-trainCarLabeling-two}
\end{figure}

The train car labeling user interface \cref{fig:dashboard-trainCarLabeling} and \cref{fig:dashboard-trainCarLabeling-two} consists of three parts: a top bar with the logo, the back button and the logout button, a left panel listing unlabeled workshop visits, and a main labeling view. The main labeling view features input fields for recording train car entries, exits, and identification numbers. This layout ensures a straightforward and organized workflow for annotators. Below, the labels for the found faults and repaired faults are listed as selectable items. 

\begin{figure}
    \centering
    \includegraphics[width=1\linewidth]{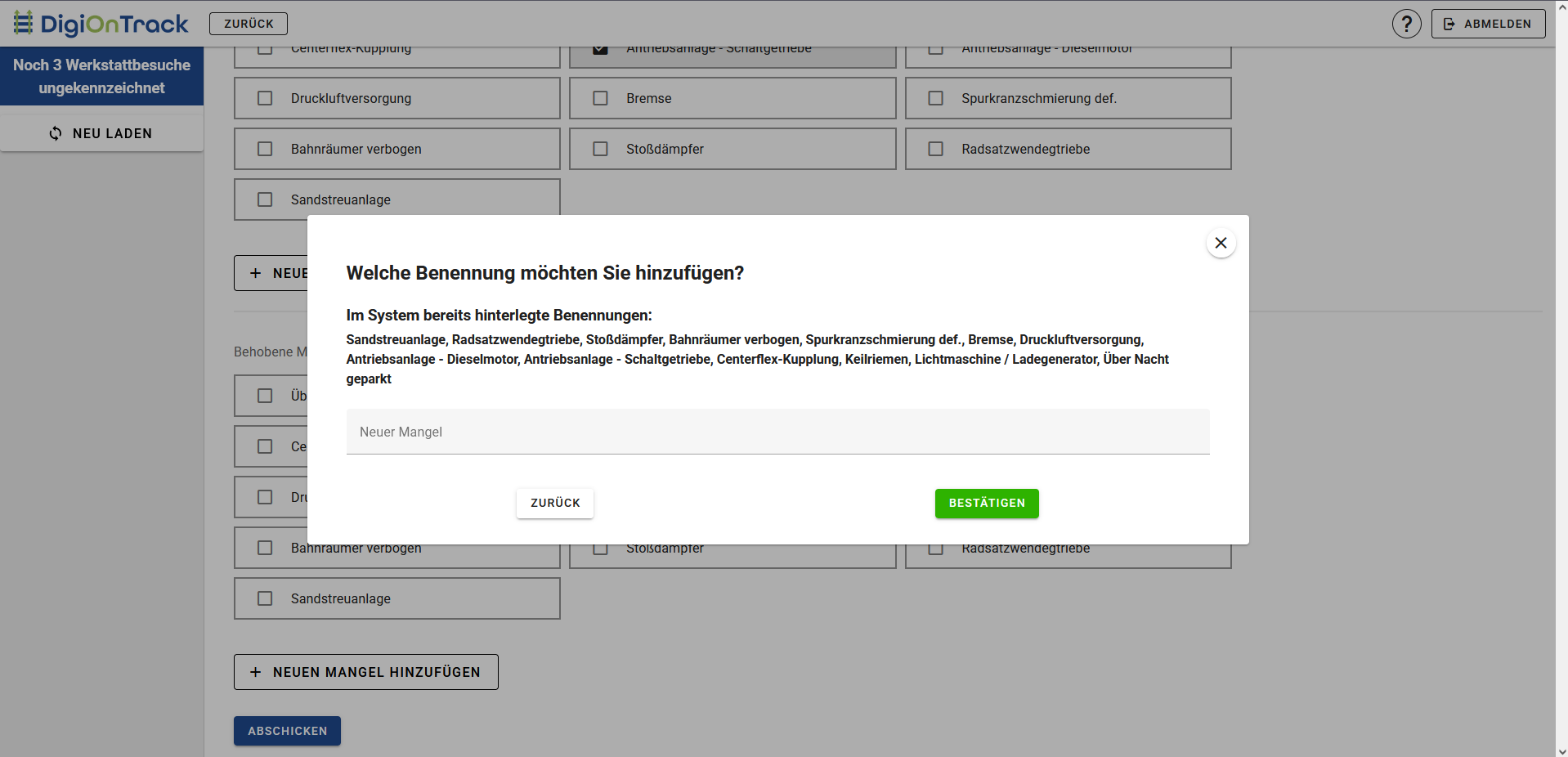}
    \caption{Overlay for Creating a New Label: In the Center are the List with Already Available Labels and an Input Field for the New Label, at the Bottom of the Overlay are the Buttons to Go Back and to Confirm the Input, it is Possible to Close the Overlay with the Close Button in the Upper Right Corner}
    \label{fig:create-label}
\end{figure}

No label hierarchy was implemented to maintain a minimalist interface and avoid lengthy interactions. The workshop foreman can create new labels via a button below the label lists. The system supports selecting and creating multiple labels for labeling multiple faults. When the workshop foreman clicks on "Neuen Mangel Hinzufügen" (eng. "create new label"), a simple overlay appears with an input field for the label name \cref{fig:create-label}. Above the input field, all existing labels are displayed, providing an overview and preventing duplicate label creation, aligning with the "Error Prevention" heuristic \cite{ref_UsabilityEngineering}. The overlay includes close, back, and confirm buttons. The close button enhances "Consistency and Standards," while the close and back buttons ensure "User Control and Freedom," allowing easy return to the labeling interface \cite{ref_UsabilityEngineering}.

\begin{figure}
    \centering
    \includegraphics[width=1\linewidth]{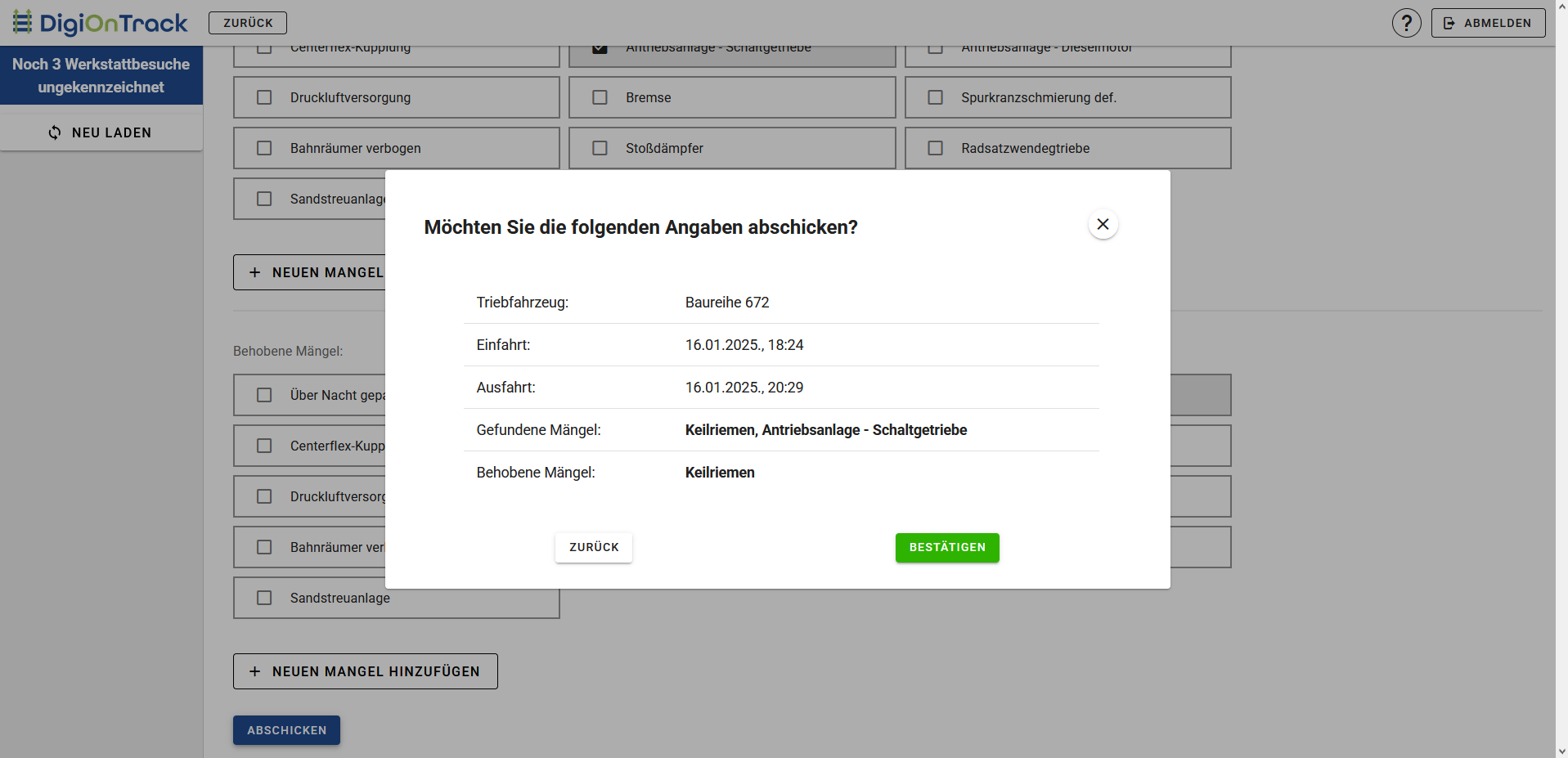}
    \caption{Data Verification Before Submitting Labels: User is Asked if the Following Data Should Be Submitted, Train Car Data with Selected Labels is Listed, at the Bottom of the Overlay are the Buttons to Go Back and to Confirm the Input, it is Possible to Close the Overlay with the Close Button in the Upper Right Corner}
    \label{fig:data-verification}
\end{figure}

When the workshop foreman selects labels and clicks the blue send button, an overlay appears with all input data for data verification \cref{fig:data-verification}. This allows corrections, ensures system status visibility by displaying saved data, and aligns with Usability heuristics \cite{ref_UsabilityEngineering}. The minimalist overlay includes close, back, and confirm buttons, giving users full control and flexibility to close, return to the labeling dashboard, or submit the data.

\subsubsection{Dashboard for Rail Faults Labeling}

\begin{figure}
    \centering
    \includegraphics[width=0.5\linewidth]{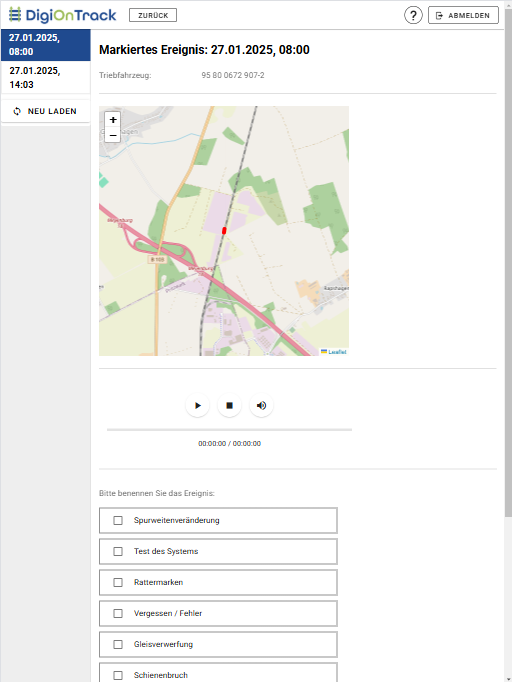}
    \caption{Dashboard for Rail Faults Labeling of the Train Car Drivers: List of Events on the Left, Date, Time, Train Identification and Event Location are shown at the Top, Label List at the Bottom}
    \label{fig:rails-labeling}
\end{figure}

\begin{figure}
    \centering
    \includegraphics[width=0.5\linewidth]{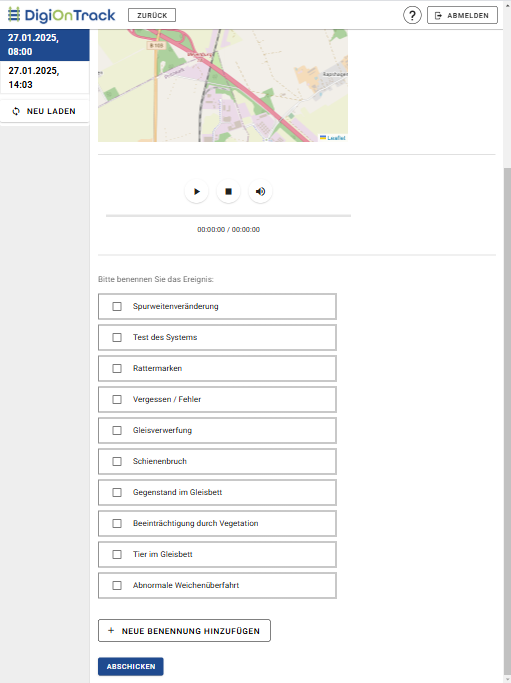}
    \caption{Dashboard for Rail Faults Labeling of the Train Car Drivers - Continuation: Submit Button is at the Bottom}
    \label{fig:rails-labeling-02}
\end{figure}

The rail labeling user interface \cref{fig:rails-labeling} \cref{fig:rails-labeling-02}, designed for train drivers' tablets, mirrors the train car labeling user interface. Events with their date and time are listed on the left, while the top features a logo, a back button, and a logout button, ensuring recognition and user control \cite{ref_UsabilityEngineering}.

The main screen displays the selected event's details—date, time, train identification number, and location—providing system status visibility \cite{ref_UsabilityEngineering}. Location is shown on a map using OpenStreetMap \cite{ref_OpenStreetMap}, with a polyline marking where the event was tagged during the journey. Drivers can zoom and navigate the map by touch for better orientation.

Labels follow a flat hierarchy, but a hierarchical structure may be introduced for improved organization if the list grows too large.

\section{Methods}
\subsection{Study Design}

A task-based user study was conducted to evaluate the Usability and User Experience of the labeling system. The study focused on both user roles in our system, the locomotive drivers and the workshop foreman. Participants were asked to perform two labeling tasks for each role, resulting in two user scenarios. The tasks per user scenario were limited to two, as the labeling in the field shows hardly any variation, and the system is very heterogeneous. Various tools of the labeling system were included in the tasks. The user had to navigate through the start page to his part of the system, select the train he was driving as a train driver, and create and select a new label—the study design aimed to measure the labeling process's ease and efficiency while identifying areas for improvement.

\subsection{Participant Selection and Ethical Considerations}

We recruited five design experts from our university who evaluated the labeling system in our Usability study and gave us additional qualitative feedback. All participants have a design-related background and are familiar with digital systems. All participants provided their informed consent. The local ethics committee approved the study.

\subsection{Test Procedure}

At the beginning of the test, the participant received a detailed explanation of the research project's purpose and use of the labeling user interface. After the participant had no more queries and answered the questionnaire's demographic questions, the two labeling tasks for the locomotive driver scenario were given by text. The participant performed the tasks on a tablet, as the locomotive drivers use the labeling system on their company's tablet. After performing the tasks, the participant was given the User Experience Questionnaire and the System Usability Scale to evaluate this part of the system. Next were the two labeling tasks for the workshop foreman, equally given to the participant by text. Participants performed the tasks on a computer, as the workshop foreman uses a computer for the labeling in his workshop office. After performing the tasks, the participant again filled out the User Experience Questionnaire and System Usability Scale to evaluate this part of the system.

\subsection{Measures}

Measures in this study, including the System Usability Scale (SUS) and the User Experience Questionnaire (UEQ), are described in the following sections.

\subsubsection{SUS Overview}
The SUS is the most widely used standardized scale to measure the perceived Usability of a System \cite{ref_SUSPaper} and thus has been used to measure the Usability of the Labeling System in this study. The SUS consists of a 5-point Likert Scale ranging from "Strongly Disagree" to "Strongly Agree" and consists of 10 items. The metric consists of scores ranging from 0 to 100, where higher scores indicate better Usability. A score above 68 is generally considered above average Usability \cite{ref_SUS, Lewis_SUS}. The content of the items asks, for example, about the system's complexity, learnability, and consistency.

\subsubsection{UEQ Overview}

Though the SUS is a proven measure for Usability, a key aspect of User Experience, it is insufficient to measure the broader concept of User Experience \cite{Lewis_SUS}. Therefore, we included the User Experience Questionnaire, a standardized and reliable measure of a system's perceived User Experience through subjective quality ratings like Attractiveness, Ffficiency, and Dependability \cite{ref_UEQPaper}. The scale consists of 26 items rated on a 7-point scale, with opposing descriptors such as 'creative' and 'dull' anchoring the endpoints.

\subsubsection{Qualitative Assessment}
An open-ended question was included at the end of the questionnaire to gather qualitative feedback from participants about the system.

\subsection{Data Analysis}

For this study, five participants (n = 5), all design experts, were recruited to evaluate the two user interfaces developed for locomotive drivers and workshop foreman. Among the participants, 3 (60\%) identified as male, and 2 (40\%) identified as female. Their ages ranged from 27 to 35 years, with a mean age of 31.00 (SD = 3.27).

\textbf{Data Calculation and Analysis}

The System Usability Scale (SUS) scores were calculated using the standard scoring method, where responses to positively worded items were adjusted as (X-1), and responses to negatively worded items were adjusted as (5-X). The total score was then multiplied by 2.5 to generate the final SUS score for each participant. For the User Experience Questionnaire (UEQ), mean scores and standard deviations were computed for each of the six dimensions: Attractiveness, Perspicuity, Efficiency, Dependability, Stimulation, and Novelty. Additionally, an overall UEQ score was calculated for each condition by averaging the scores across all six dimensions to provide a general measure of User Experience.

\newpage\textbf{Statistical Analysis}

To summarize the data, descriptive statistics (mean and standard deviation) were computed for all SUS scores, UEQ dimensions, and the overall UEQ score to summarize the data. To assess whether the differences between the two interfaces for each dimension, the overall UEQ score, and the SUS scores were statistically significant, a paired samples t-test was conducted. A significance threshold of $p \leq 0.05$ was used to identify meaningful differences.

\section{Results}

\subsection{SUS Results}

\begin{figure}[H]
    \centering
    \includegraphics[width=0.8\textwidth]{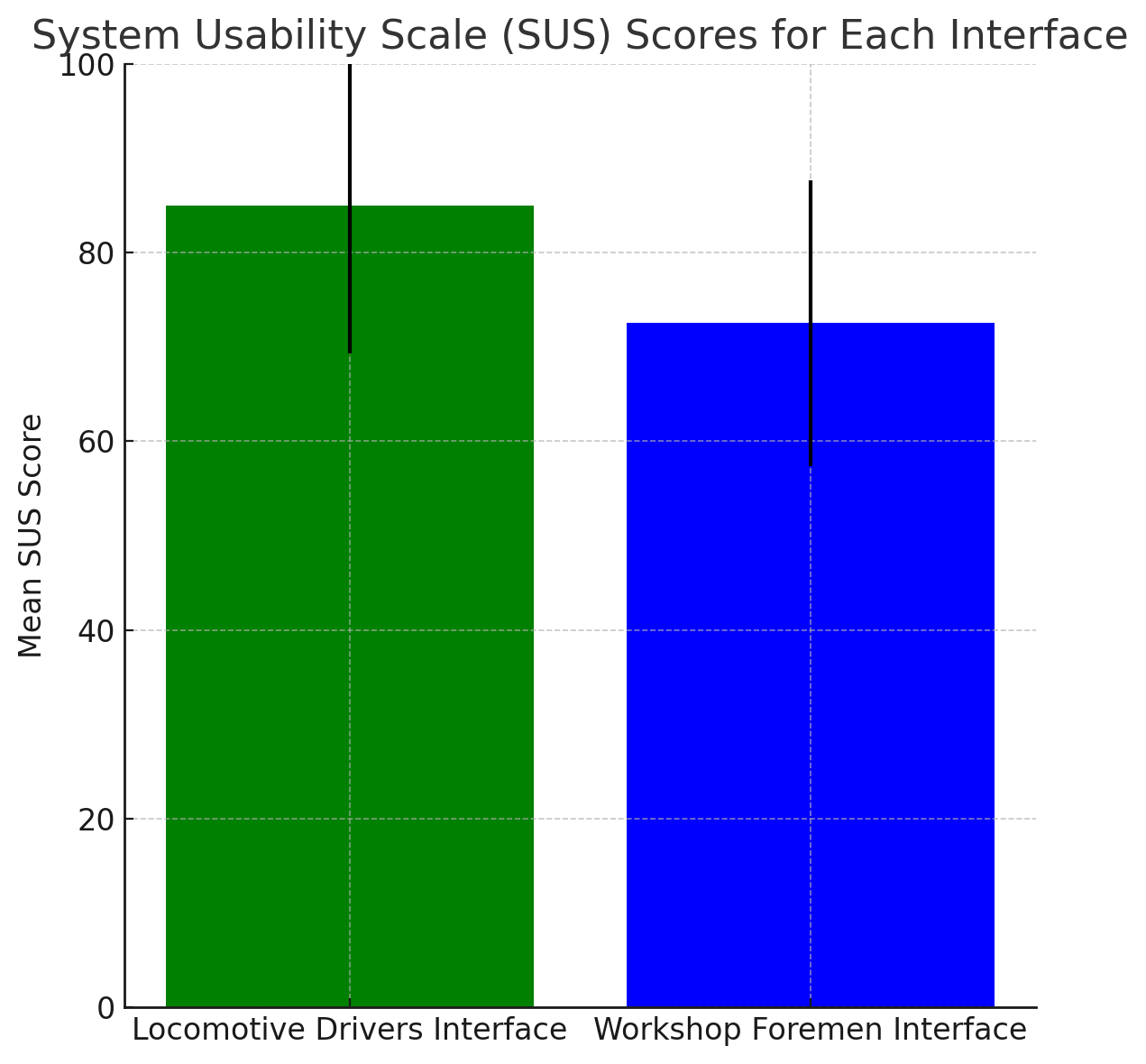} 
    \caption{Mean SUS scores for the two evaluated interfaces (locomotive drivers and workshop foreman), as rated by design experts. Locomotive drivers’ interface: $M = 85.00$, $SD = 15.71$; Workshop foreman’s interface: $M = 72.50$, $SD = 15.10$. Error bars represent the standard error of the mean. A statistically significant difference between the two interfaces was observed ($p = .035$, * = $p \leq 0.05$).}
    \label{fig:sus_scores}
\end{figure}

The mean SUS score, see \cref{fig:sus_scores}, for the locomotive drivers’ interface was $M = 85.00$ ($SD = 15.71$). In contrast, the workshop foreman’s interface achieved a lower mean score of $M = 72.50$ ($SD = 15.10$). The difference between the two interface scores was statistically significant ($p = .035, t(39) = 2.15$).

\subsection{UEQ Results}

\begin{figure}[H]
    \centering
    \includegraphics[width=0.8\textwidth]{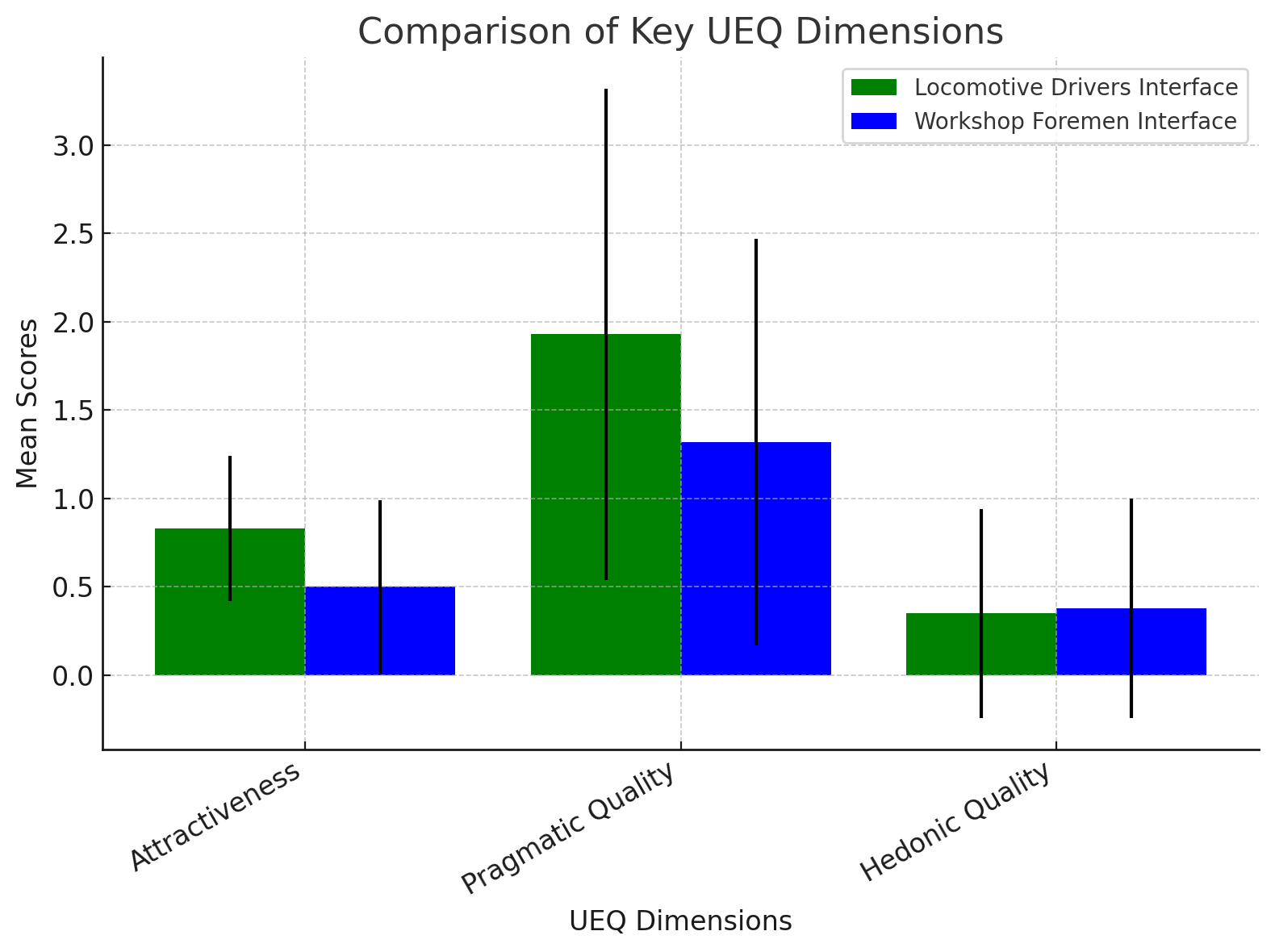} 
    \caption{Mean scores for the three key UEQ dimensions (Attractiveness, Pragmatic Quality, and Hedonic Quality) for the two evaluated interfaces (locomotive drivers and workshop foreman). The locomotive drivers’ interface achieved the following scores: Attractiveness ($M=0.83, SD=0.41$), Pragmatic Quality ($M=1.93, SD=1.39$), and Hedonic Quality (M=0.35, SD=0.59). The workshop foreman’s interface was rated as follows: Attractiveness ($M=0.50, SD=0.49$), Pragmatic Quality ($M=1.32, SD=1.15$), and Hedonic Quality ($M=0.38, SD=0.62$). Error bars represent the standard deviation. No statistically significant differences were found between the interfaces across these dimensions.}
    \label{fig:ueq_scores}
\end{figure}

The interface designed for \textbf{locomotive drivers} achieved in the UEQ, see \cref{fig:ueq_scores}, a mean score of $M = 1.04$ ($SD = 0.76$), while the \textbf{workshop foreman’s} interface received a mean score of $M = 0.73$ ($SD = 0.58$). 
\begin{table}[H]
    \centering
    \caption{Overall UEQ Score}
    \begin{tabular}{lcc}
        \toprule
        \textbf{Interface} & \textbf{Mean} & \textbf{Standard Deviation} \\
        \midrule
        Locomotive Drivers Interface & 1.04 & 0.76 \\
        Workshop foreman Interface & 0.73 & 0.58 \\
        \bottomrule
    \end{tabular}
    \label{tab:ueq_overall}
\end{table}

A detailed breakdown of the three key dimensions provides further insights. \textbf{Attractiveness}, representing the overall appeal of the system, was rated higher for the locomotive drivers’ interface ($M = 0.83, SD = 0.41$) than for the workshop foreman’s interface ($M = 0.50, SD = 0.49$). \textbf{Pragmatic Quality}, which encompasses Efficiency, Perspicuity, and Dependability, also showed a higher rating for the locomotive drivers’ interface ($M = 1.93, SD = 1.39$) compared to the workshop foreman’s interface ($M = 1.32, SD = 1.15$). \textbf{Hedonic Quality}, related to Stimulation and Novelty, was nearly identical across both interfaces, with the locomotive drivers' interface scoring $M = 0.35$ ($SD = 0.59$) and the workshop foreman’s interface $M = 0.38$ ($SD = 0.62$).

A further breakdown into the six UEQ dimensions reveals more details, also shown in \cref{tab:ueq_results}. \textbf{Attractiveness} was perceived as higher for the locomotive drivers’ interface ($M = 0.83, SD = 0.41$) compared to the workshop foreman’s interface ($M = 0.50, SD = 0.49$). \\

\textbf{Perspicuity}, which reflects ease of learning, was rated higher for the locomotive drivers' interface ($M = 2.00, SD = 0.94$) than for the workshop foreman’s interface ($M = 1.05, SD = 1.39$).\\

In terms of \textbf{Efficiency}, which measures how quickly and effectively users could complete tasks, the locomotive drivers’ interface ($M = 2.25, SD = 0.92$) outperformed the workshop foreman’s interface ($M = 1.65, SD = 0.96$). \\

\textbf{Dependability}, which evaluates reliability and predictability, was rated at ($M = 1.55$, $SD = 0.76$) for the locomotive drivers’ interface and slightly lower at ($M = 1.25$, $SD = 0.56$) for the workshop foreman’s interface.\\

\textbf{Stimulation}, assessing how engaging and motivating an interface is, was slightly lower for the locomotive drivers’ interface ($M = 0.70, SD = 0.48$) compared to the workshop foreman’s interface ($M = 0.90, SD = 1.09$). \\

Lastly, \textbf{Novelty}, which reflects the innovativeness of the interface, was rated lowest among all dimensions, with the locomotive drivers’ interface scoring ($M = 0.00$ ($SD = 1.34$) and the workshop foreman’s interface slightly lower at ($M = -0.15$, $SD = 1.17$).\\

\begin{table}[H]
    \centering
    \caption{Mean and Standard Deviation of UEQ Dimensions}
    \begin{tabular}{lcc|cc}
        \toprule
        \textbf{UEQ Dimension} & \multicolumn{2}{c}{\textbf{Locomotive Driver's Interface}} & \multicolumn{2}{c}{\textbf{Workshop foreman Interface}} \\
        & Mean & SD & Mean & SD \\
        \midrule
        Attractiveness & 0.83 & 0.41 & 0.50 & 0.49 \\
        Perspicuity    & 2.00 & 0.94 & 1.05 & 1.39 \\
        Efficiency     & 2.25 & 0.92 & 1.65 & 0.96 \\
        Dependability  & 1.55 & 0.76 & 1.25 & 0.56 \\
        Stimulation    & 0.70 & 0.48 & 0.90 & 1.09 \\
        Novelty        & 0.00 & 1.34 & -0.15 & 1.17 \\
        \bottomrule
    \end{tabular}
    \label{tab:ueq_results}
\end{table}

\subsection{Qualitative Assessment Results}
In the qualitative survey, one participant criticized that entering different entry and exit times can quickly become “confusing”. Another participant pointed out that it is "impossible to edit the data sent back into the system", making using the system prone to errors.

\section{Discussion}

In this study, we evaluated the Usability and User Experience of the labeling system designed and developed for our research project 'DigiOnTrack.' This project assesses the effectiveness of combining structure-borne noise measurement methods with supervised machine learning to provide monitoring and maintenance recommendations for rail vehicles and infrastructure.

The locomotive drivers’ interface achieved higher Usability and User Experience ratings compared to the workshop foreman’s interface, with significant differences in overall SUS scores and advantages across the key dimensions of Attractiveness, Pragmatic Quality, and the sub-dimensions Perspicuity, Efficiency, Dependability, and Novelty. The mean SUS score for the locomotive driver's interface can be categorized as "Excellent Usability"  \cite{ref_SUSPaper}, while the workshop foreman’s interface falls within the range of "Good Usability". The benchmark of the official UEQ-evaluation tool \cite{ref_UEQ} classifies the locomotive drivers' interface's Attractiveness as "Below Average", its Perspicuity as "Good", its Efficieny as "Excellent", its Dependability as "Good", its Stimulation as "Below Average" and its Novelty as "Bad". Meanwhile, the benchmark classifies the workshop foreman's interface's Attractiveness as "Bad", its Perspicuity as "Below Average", its Efficiency as "Good", its Dependability as "Above Average", its Stimulation as "Below Average" and its Novelty as "Bad".

In this study, we evaluated two labeling user interfaces whose designs focused on minimalism, simplicity, and learnability to make labeling as easy and fast as possible for the annotators. Looking at the related dimensions of the UEQ, i.e., Perspicuity, Efficiency, and Dependability, it can be suggested that the interface for the train driver has been successfully designed and implemented. The categorization of "Excellent Usability" of the SUS emphasizes this conclusion. The interface for the workshop foreman is classified as "Good" in Efficiency and scores "Above Average" in Dependability, but has a "Below Average" classification in Perspicuity. This indicates that the system contributes to a high level of labeling efficiency but is slightly more challenging to learn and operate. This could be due to the nature of the dashboard, where due to data lack, as in many predictive maintenance research projects in the railway domain \cite{Binder}, the workshop foreman has to add a lot of data in addition to the labels with entry, exit and train identification number. One participant's statement in the qualitative assessment emphasizes this indication that the input of entry and exit by listing by times makes the system confusing. The "Below Average" results for the dimensions “Stimulation” and “Novelty” in both user interfaces could be due to the nature of the use case, as labeling faults can be considered tedious and annoying \cite{ref_TediousLabeling}. However, as the workshop foreman's interface is classified as "Good Usability", the system seems to have been implemented well and supportively in accordance with the circumstances.
Although the study provides valuable insights into the system's Usability and User Experience through the evaluation by design experts, it is important to note that the participant pool was limited and did not include individuals from the annotator groups. A more extensive study involving non-design experts could be conducted to achieve a more generalized system evaluation. Additionally, a field test with participants from the annotator groups would be particularly valuable for performing a user-centered evaluation, offering more profound insights into its practical applicability. If multiple assessments of different labeling systems were available, these could serve as comparisons for newly developed ones, enabling a more focused and comprehensive evaluation. Additionally, further published studies would provide valuable insights to guide decision-making in the design of labeling systems. Ultimately, drafting guidelines based on these findings could assist future designers and developers create effective labeling systems, thereby contributing to the successful implementation of predictive maintenance systems powered by supervised machine learning algorithms.

\section{Conclusion}
We presented a user-centered labeling and data management system for the cost-effective implementation of predictive maintenance for rail infrastructure and vehicles in rural Germany. We developed a system that integrates wireless sensor networks, data analytics, and a secure data transfer to monitor the condition of rail infrastructure and vehicles and predict maintenance tasks in the future. A critical component was the custom-built labeling system, which served to annotate data for supervised machine learning algorithms, promoting accurate predictive maintenance recommendations. The Usability and User Experience evaluations highlighted that the locomotive drivers’ interface achieved "Excellent Usability." In contrast, the workshop foreman’s interface was rated as "Good Usability", indicating that the labeling system can be integrated seamlessly into the annotators' daily work routines. Further, the critical dimension for the labeling use case, "Efficiency", of the User Experience Questionnaire shows a "Good" to "Excellent" classification for the two interfaces, showing the efficiency of the system. While the interface for the workshop foreman lacks in "Perspicuity", probably through the lack of data for the dashboard and the resulting conspicuousness of the interface, it indicates that further optimization is needed to enhance Usability and learnability in more data-intensive scenarios. Future studies could build on these findings by conducting more extensive evaluations with non-design experts and field tests involving annotator groups to achieve a more comprehensive and user-centered system validation. Comparative assessment with other labeling systems could also provide deeper insights into best practices for designing effective interfaces. Ultimately, the results from this work could inform the development of guidelines for creating efficient and user-friendly labeling systems, not only in predictive maintenance but also in broader Industry 4.0 applications.

\begin{credits}
\subsubsection{\ackname} We gratefully acknowledge financial support through the TÜV Rheinland Consulting GmbH with funds provided by the Federal Ministry for Digital and Transport (BMDV) under Grant No. 19F2265 (DigiOnTrack). We also want to thank the reviewers for their thoughtful feedback.

In this paper, we used Overleaf’s built-in spell checker, the current version of ChatGPT (GPT 3.5), and Grammarly. These tools helped us fix spelling mistakes and get suggestions to improve our writing. If not noted otherwise in a specific section, these tools were not used in other forms.

\subsubsection{\discintname}
The authors have no competing interests to declare relevant to this article's content. 
\end{credits}
%
%
%
 \bibliographystyle{splncs04}
 \bibliography{main}

\end{document}